\documentclass[prb, showpacs, twocolumn]{revtex4}

\usepackage{graphicx}
\usepackage{amsmath}
\usepackage{amsbsy}
\usepackage{amssymb}

\usepackage{array}

\DeclareGraphicsExtensions{.eps,.pdf} 

\begin{document}

\title{Resonant spin polarization and spin current in a two-dimensional
  electron gas}

\author{Mathias Duckheim}

\author{Daniel Loss}

\affiliation{Department of Physics and Astronomy, University of Basel, CH-4056
Basel, Switzerland}
\date{\today}


\begin{abstract}
  We study the spin polarization and its associated spin-Hall current
  due to electric-dipole-induced spin resonance in disordered
  two-dimensional electron systems. We show that the disorder induced
  damping of the resonant spin polarization can be strongly reduced by
  an optimal field configuration that exploits the interference
  between Rashba and Dresselhaus spin-orbit interaction. This leads to
  a striking enhancement of the spin susceptibility while the
  spin-Hall current vanishes at the same time. We give an
  interpretation of the spin current in geometrical terms which are
  associated with the trajectories the polarization describes in spin
  space.
\end{abstract}

\maketitle

The ability to coherently control the spin of charge carriers in semiconductor
nanostructures is the main focus of spintronics\cite{spintronics}.
Band-structure and confinement effects in these systems lead to a strong
spin-orbit interaction (SOI) offering the possibility to efficiently access
the charge carrier spin via the control of its orbital motion\cite{edelstein,
  kato, Rashba2003a, dyakonov, Murakami2004, Sinova2004, awschalom,
  awschalom-she-2deg, Bulaev2007, Golovach2006, Trif2007, Shnirman2007}.

A versatile and efficient scheme of spin control is electric dipole induced
spin resonance (EDSR)\cite{Bell1962, Dobrowolska1984, Rashba2003, Schulte2005,
  Kato2004, Golovach2006, Bulaev2007, Duckheim2006, Wilamowski2007} where
electric radio frequency (rf) fields give rise to internal fields coupling to
the spin. Choosing an adequate configuration of the electric rf fields and a
static magnetic field defining a quantization axis for the spin, arbitrary
spin rotations can be realized. This is analogous to standard paramagnetic
spin resonance techniques, has the advantage, however, that it can be
integrated in gated nanostructures thereby avoiding magnetic rf coils.

\begin{figure}[t]
  \centering
\includegraphics[width = 8.5cm]{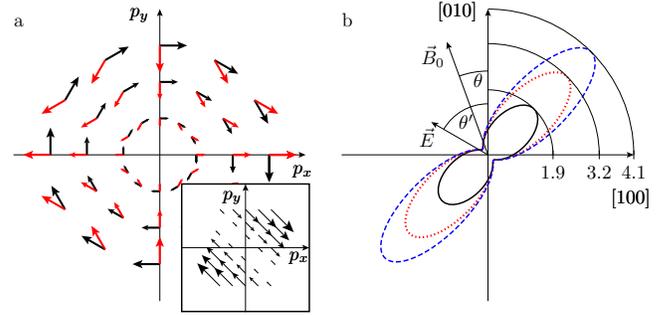}  
\caption{(color online) a) Momentum dependent magnetic fields induced by
  Rashba- (black arrows) and Dresselhaus SOI (red arrows) $\mathbf
  \Omega_R(\mathbf p) = \alpha (p_y, -p_x)$ and $\mathbf \Omega_D(\mathbf p) =
  \beta (p_x, -p_y)$, respectively. Inset: The sum $\mathbf \Omega_R + \mathbf
  \Omega_D$ for equal strength of the SOI ($\alpha = \beta$) is shown. The
  interference of the two types of SOI leads to a suppression or enhancement
  of the spin splitting in certain crystallographic directions.  b) Polar plot
  of the resonance susceptibility $\bar \chi^{\mathrm{res}}$ (in arbitrary
  units) as a function of $\theta$ for $ \beta = \alpha/ 2$ and $\omega_L \tau
  = 1$ (black, solid curve), $\omega_L \tau = 2$(red, dotted), and $\omega_L
  \tau = 3$(blue, dashed).  The configuration of the external magnetic and electric field
  $\mathbf B_0$ and $\mathbf E_0$ is shown.  For $\mathbf B_0 || \mathbf E_0
  || [110]$ both SOI contributions add constructively in the direction
  perpendicular to $\mathbf B_0$ leading to an enlarged Rabi field.
}
  \label{fig:vectorfields}
\end{figure}

In a two-dimensional electron gas (2DEG) with pure Rashba SOI the amount of
spin polarization which can be achieved by EDSR is severely limited by
disorder\cite{Duckheim2006}. Similar limitations are found for pure
Dresselhaus SOI. However, if both Dresselhaus and Rashba SOI are present
interference between the two SOI mechanisms can occur and qualitatively new
behavior emerges, such as anisotropy in spin relaxation\cite{Averkiev2002,
  Golovach2004, Li2006} and transport\cite{Schliemann2003a, Ganichev2004,
  Trushin2007}. For spin relaxation this anisotropy is most pronounced if both
SOIs have equal strength.  In this case, the spin along the $[1 1 0]$
direction [see Fig.  \ref{fig:vectorfields}] is conserved\cite{Schliemann2003,
  Averkiev2002}, and the associated spin relaxation rates vanish, whereas they
become maximal along the perpendicular direction $[1\bar 1 0]$.  For the
driven system considered here we show that similar interference effects occur
and that not only the internal rf field but also the EDSR linewidth becomes
dependent on the direction of the magnetic field.  In a microscopic approach
we show then that due to this dependence an optimal configuration exists where
the linewidth and the internal field simultaneously become minimal and
maximal, resp., and that, as a remarkable consequence, the spin susceptibility
gets dramatically enhanced.  In other words, this optimal configuration allows
one to obtain a high spin polarization with relatively small electric fields
and thus making the power consumption for spin polarization minimal.

Due to spin-orbit interaction angular momentum can be transferred between spin
and orbital degrees of freedom. This fact leads, in particular, to a dynamical
coupling between spin and spin current described by the Heisenberg equation of
motion\cite{Erlingsson2005, Chalaev2005, Duckheim2006}.  Exploiting this
coupling we show that the spin current can be interpreted in geometrical
terms: the spin dynamics generated by the rf fields describes an elliptical
trajectory. The spin-Hall conductivity can then be expressed entirely in terms
of the semi-minor and semi-major axis and the tilt angle of this ellipse.
Since the spin dynamics (trajectories) is experimentally accessible, for
instance with optical methods\cite{awschalom, awschalom-she-2deg}, this opens
up the possibility for a direct measurement of the spin-Hall current.  Finally,
we find that for the optimal configuration the spin current vanishes, in stark
contrast to the spin polarization which, as mentioned, becomes maximal.

We consider a non-interacting 2DEG consisting of electrons with mass
$m$ and charge $e$ which are subject to a random impurity potential
$V$.  We take into account linear SOI $\sum_{ij} \mathbf \Omega_{ij}
p_j \sigma^i$ of the Rashba - and Dresselhaus type where $\sigma^i,
i=1,2,3,$ are the Pauli matrices and $\mathbf p$ is the canonical
momentum. Taking the coordinate axes along the $[100], [010], [001]$
crystallographic directions, the internal magnetic field $\mathbf
\Omega$ is then given by (cf.  Fig.\ref{fig:vectorfields}) $\mathbf
\Omega(\mathbf p) = \alpha (p_y, -p_x,0) + \beta (p_x, -p_y,0)$ where
$\alpha$ and $\beta$ is the strength of the Rashba and Dresselhaus
SOI, respectively. Additionally, the external static magnetic field is
given by $\mathbf B_0 = B_0 \mathbf e_{||}$ with $ \mathbf e_{||} =(-
\sin \theta, \cos \theta,0)$, and the external electric rf field by
$\mathbf E(t) = E(t) (- \sin \theta', \cos \theta',0)$, where $\theta,
\theta'$ are the angles enclosed with the $[010]$ direction.  The
system is described by the Hamiltonian
\begin{equation}
  \label{eq:h_ext}
  H = \frac{1}{2m} \left( \mathbf p - e \mathbf A \right)^2 +
  \left( \mathbf \Omega \left( \mathbf p - e \mathbf A \right) +
    \mathbf b_0 \right) \cdot
  \boldsymbol{\sigma} + V,  
\end{equation}
where $\mathbf A(t) = - \int^t dt' \mathbf E(t')$ is the vector potential
associated with $\mathbf E$ and $\boldsymbol{b}_0 = g \mu_B \mathbf B_0 /2$
with $g$ the electron g-factor and $\mu_B$ the Bohr magneton.

{\it Spin polarization.}  We turn now to the calculation of the spin
polarization (magnetization$/ \mu_B$) per unit area, $\mathbf S(\omega) =
\int_{-\infty}^{\infty} dt e^{i \omega t} \langle \boldsymbol{\sigma} (t)
\rangle/2 \pi $, evaluated in linear response to an applied electric field
$\mathbf E(\omega) = \mathbf E_0 [\delta(\omega - \omega_0) + \delta(\omega +
\omega_0)]/2$ and in the presence of both Rashba and Dresselhaus SOI.  Due to
the interference between these two SOI mechanisms we need to carefully extend
earlier calculations\cite{Duckheim2006}, which were restricted to Rashba
(Dresselhaus) SOI only, to this new situation. We will then be able to
identify a configuration that allows one to obtain a maximum degree of spin
polarization due to this interference.

Working in the linear response regime, $S^i(\omega)$ is obtained from a Kubo
formula averaged over the random distribution of impurities in the 2DEG. We
evaluate this average with standard diagrammatic techniques assuming the
impurities to be short-ranged, isotropic and uniformly distributed. In this
case, the impurity average $\overline{V(\mathbf x)V(\mathbf x')} \equiv (m
\tau)^{-1} \delta(\mathbf x-\mathbf x')$ is $\delta$ - correlated and
proportional to the momentum relaxation time $\tau$. We further take the Fermi
energy $E_F = p_F^2/2m$ to be the largest energy scale in the problem.  Then,
to leading order in $1/p_F l$ with $l = p_F/m \tau$ the mean free path, the
averaged spin polarization is given by the diffuson diagram, giving rise
\cite{rammer} to a correction $\sigma^i \rightarrow  \Sigma^i \equiv 
\Sigma^{ij} \sigma^j$ of the spin vertex (cf.  ref.  \cite{Duckheim2006}) in
the Kubo formula. Thus, the spin susceptibility defined by $S^i(\omega) =
\chi^{ij}(\omega) E_j(\omega) $ is given by
\begin{align}
  \label{eq:spin-pol-rd}
  \chi^{ij} (\omega) = \sum_{k=1}^{k=3} 2 e \nu \tau \Bigg[ \delta^{i k} - \left(1-
    \frac{1}{\lambda} \right) \Sigma^{i k} \Bigg] \Omega_{kj}  \, ,
\end{align}
where $\nu = m/2 \pi \hbar^2$ is the two-dimensional density of states
and $\lambda(\omega) = 1 - i \omega \tau$.

We evaluate the vertex correction $\Sigma$ of
Eq.(\ref{eq:spin-pol-rd}) for the case of a magnetic field $B_0$ that
is large compared to the internal fields induced by SOI. This regime
is characterized by $a_R \equiv \alpha p_F/2 b_0 \ll 1$ and $a_D
\equiv \beta p_F/2 b_0 \ll 1$. The components $\Sigma^{ij}$ with $i,j
=1,2,3$ of the vertex correction are then found to be given by
\begin{widetext}
\begin{align}
  \label{eq:vertex-correction}
  \Sigma  = \frac{1}{(\omega_L^2 - \omega^2) \tau^2 + q} 
   \left(
    \begin{array}{ccc}
      y^2 \frac{\lambda - \cos^2(\theta)}{\lambda -1}+ \lambda (\lambda -1)  +
      q_{11} &  \frac{- y^2 }{2 (\lambda -1)} \sin (2 \theta) +  q_{12} & y
      \cos (\theta) + q_{13}\\
      \frac{- y^2 }{2 (\lambda -1) } \sin (2 \theta)  + q_{12} &     y^2
      \frac{\lambda - \sin^2(\theta)}{\lambda -1}+ \lambda (\lambda -1) + q_{22}  & y
      \sin (\theta)  + q_{23} \\  - y
      \cos (\theta) - q_{13} & - y
      \sin (\theta) - q_{23}   & y^2 + \lambda (\lambda -1) + q_{33} 
    \end{array}
  \right)   \, ,
\end{align}
\end{widetext}
where $y=2 b_0 \tau/\hbar = \omega_L \tau$.  Here, the functions $q_{ij}$ and
$q$ are second order in $a_R$ and $a_D$, and depend on the frequency $\omega$,
the Larmor frequency $\omega_L$, and the angle $\theta$\footnote{The $q_{ij}$
  and $q$ are too lengthy to be written down here. In the following, only
  $\mathrm{Im} \, q$, $\mathrm{Re} \, q$ and the linear combination $\sqrt{2}
  (q_{13} - q_{23}) y + (q_{11} - 2 q_{12} + q_{22} )(\lambda -1) -2 \lambda
  q$ are relevant for the linewidth and the spin-Hall conductivity, resp.,
  which are explicitly given below in Eqs.(\ref{eq:damping2}) and
  (\ref{eq:spin-current-explicit}).}. In the EDSR system, Pauli paramagnetism
gives rise to a constant equilibrium polarization $\mathbf S_{\mathrm{eq}} =
\nu \hbar \omega_L \mathbf e_{||}$ along $\mathbf B_0$ which is independent of
the electric field. The polarization dynamically generated by $\mathbf
E(\omega)$, however, depends on the amplitude of the oscillating internal
field perpendicular to $\mathbf B_0$.  It is thus instructive to consider the
longitudinal (along $\mathbf B_0 || \mathbf e_{||}$) and the transverse (along
$\mathbf e_{\perp} \equiv \mathbf e_{||} \times \mathbf e_3$ and $\mathbf
e_3$) polarization components given by $S'^2 = \mathbf{e}_{||} \cdot \mathbf
S$ and $S'^1 = \mathbf{e}_\perp \cdot \mathbf S$, $S'^3 = S^3$, resp.

As a result, we find the polarization $S'^i(\omega) = \bar \chi^{i}(\omega,
\theta') E(\omega) $ in terms of the transformed susceptibility $\bar \chi$.
To lowest order in $a_R, a_D$ only the transverse components ($i = 1,3$) are
finite. They are given by
\begin{align}
  \label{eq:sus3trans}
  &\bar \chi^{i}( \omega, \theta') = S_{\mathrm{eq}} \, l(\omega)  \left[ \alpha \cos (\theta' - 
\theta) - \beta \sin
    (\theta' + \theta) \right] \notag \\ &\times
w_{i}
\left(
    \frac{1}{\omega_L - \omega +\delta \omega -i\Gamma} +
    \frac{1}{\omega_L + \omega -\delta \omega +i\Gamma} \right) ,
\end{align}
where $w_1 = 1$ for the in-plane ($i=1$) and $w_3 = -i
\omega/\omega_L $ for the out-of-plane component ($i=3$), and $l(\omega)
= e \tau/ \hbar (1 - i \omega \tau)$ is proportional to the Drude
conductivity\footnote{ From Eq.(\ref{eq:sus3trans}) we can identify the
  component of the internal rf field $b_1(\omega)$ (which effectively
  drives the spin dynamics) in terms of the electrically induced momentum
  drift $\langle \mathbf p \rangle (\omega)= \hbar l(\omega) \mathbf
  E(\omega)$. We find that $b_1(\omega) = \mathbf e_\perp \cdot
  \boldsymbol{\Omega} ( \langle \mathbf p \rangle (\omega))$ is given by the
  projection of the internal rf field (induced by $\langle \mathbf p \rangle
  (\omega)$) on the transverse direction $\mathbf e_\perp$.  Note that due to
  disorder scattering the Fourier transform $\langle \mathbf p \rangle
  (t)$ and $b_1(t)$ are phase-lagged with respect to $\mathbf E(t)$. }.

Close to resonance the scattering from disorder leads to a
renormalization of the magnetic field dependence. The resonance is shifted by a term
\begin{align}
  \label{eq:damping1}
  \delta \omega &= \mathrm{Re} \,q(\omega = \omega_L) /2 \omega_L
  \tau^2 \\ &=   \frac{p_F^2 \tau}{\hbar^2} [\alpha^2 + \beta^2 - 2 \beta
    \alpha \sin (2 \theta)]  \frac{ \omega_L \tau}{1 + (\omega_{L} \tau)^2}, \notag
\end{align}
corresponding to an effective g-factor which depends both on the amplitude and
the orientation of the magnetic field.  The linewidth $\Gamma$ of the
resonance peak is given by
\begin{align}
\label{eq:damping2}
\Gamma & = - \mathrm{Im} \, q(\omega = \omega_L) /2 \omega_L \tau^2 =
2 p_F^2 \tau / \hbar^2  \\ 
\times & \left[ \left( \alpha^2 + \beta^2 + 2 \alpha \beta \sin (2 \theta) \right) +
   \frac{(\alpha^2 + \beta^2) - 2 \alpha \beta \sin (2 \theta)}{ 2 [1
    + (\omega_L \tau)^2]}\right] .\notag
\end{align}
Note that in $\Gamma$ the Rashba and Dresselhaus SOI do not simply add up but
can interfere with each other, enabling a strong enhancement of the
susceptibility as we will see next. In Fig.  \ref{fig:vectorfields} we plot
the spin susceptibility at resonance, $\bar \chi^{\mathrm{res}} \propto\left[
  \cos (\theta' - \theta) - \sin (\theta' + \theta) \rho \right] /\Gamma$ for
the case $\rho = \beta / \alpha =0.5$ measured in \cite{Ganichev2004}. The
angle $\theta'$ has been tuned to maximize $\bar \chi^{\mathrm{res}}$ which
displays a pronounced dependence on the magnetic field direction. In Eq.
(\ref{eq:damping2}) we note that $\Gamma$ scales with the mean square
fluctuations of the internal magnetic fields $\langle (\mathbf e_{||} \cdot
\boldsymbol{\Omega}(p_F \hat{\mathbf{n}}))^2 \rangle_{\hat{\mathbf{n}}}$ and
$\langle (\mathbf e_\perp \cdot \boldsymbol{\Omega}(p_F \hat{\mathbf{n}}))^2
\rangle_{\hat{\mathbf{n}}}$, where $\langle .  \rangle_{\hat{\mathbf{n}}}$
denotes a uniform average over all (in-plane) directions $\hat{\mathbf{n}}$.
Comparison with a simple model\cite{Bloch1957, Tahan2005} of spin relaxation
(Bloch equation) shows that the first term in Eq.  (\ref{eq:damping2}) comes
from pure dephasing, i.e.  from disorder induced fluctuations of the internal
fields along $\mathbf B_0$, while the second term is due to fluctuations along
$\mathbf e_\perp$. Choosing a configuration with $\theta = \theta' = -\pi /4$
and tuning the SOI strengths to $\alpha = \beta$ the first term vanishes while
the second is subject to narrowing due to the magnetic field.  The width
becomes $ \Gamma_{\mathrm{DP}}/ [1 + (\omega \tau)^2]$ where $\Gamma_{DP} = 2
(\alpha p_F)^2 \tau / \hbar^2$ is the D'yakonov - Perel spin relaxation rate
for Rashba SOI.  Increasing the frequency such that (at resonance) $\omega_L
\tau = \omega \tau \gg 1 $ will lead to an increase of the inverse width
$\Gamma^{-1}$ and, hence, of the susceptibility at resonance, given by
\begin{align}
  \label{eq:pol-res}
  |\bar \chi^{\mathrm{res}}_{\alpha= \beta}| = S_{\mathrm{eq}} \frac{ e
    \alpha \tau}{\hbar \Gamma_{DP}} \sqrt{1 + (\omega_L \tau)^2}.
\end{align}
For comparison, we find the ratio to the resonance susceptibility $\bar
\chi^{\mathrm{res}}_{\beta =0}$ in the pure Rashba case as $|\bar
\chi^{\mathrm{res}}_{\alpha = \beta }/\bar \chi^{\mathrm{res}}_{\beta =0}| =
(1 + y^2) [1 + 1/(2 (1 + y^2))]$ growing quadratically with $y = \omega_L
\tau$.  Thus, the spin polarization can be substantially enhanced by tuning
the SOIs to equal strengths and by increasing the magnetic field.  Finally,
the range of validity for the linear response regime can be estimated as
follows. Assuming full polarization ($ |S'^{3,\mathrm{res}}_{\alpha=
  \beta}|/S_{\mathrm{eq}} 
\approx 1$) and parameters for a GaAs 2DEG \cite{awschalom-she-2deg} with
spin-orbit splitting $\Delta_{\mathrm{SO}} = \alpha p_F = 60\, \mathrm{\mu e
  V}$, Fermi wavelength $\lambda_F = 180 \, \mathrm{nm}$ and $\omega_L \tau =
10$, we find from Eq. (\ref{eq:pol-res}) that the linear response is valid for
electric fields with amplitudes up to $E_0 = 2 \pi \Delta_{\mathrm{SO}}/
\lambda_F \omega_L \tau \approx 200 \, \mathrm{eV} \mathrm{m}^{-1}$.

{\it Polarization and Spin current.}  We consider the spin current defined by
$\mathbf I^3 = \langle \{\sigma^3, \mathbf v \} \rangle /2$. Using the
Heisenberg equation of motion the spin current components $I^3_{x'}$ and
$I^3_{y'}$ along $\mathbf e_\perp$ and $\mathbf e_{||}$ can be expressed in
terms of the polarization at frequency $\omega$ as
\begin{align}
  \label{eq:spin-current-RD}
& \left(
  \begin{array}{c}
    I^3_{x'} \\ I^3_{y'}
  \end{array}
\right)  = \frac{\hbar}{2 m (\alpha^2 - \beta^2)} \\ & \times \left(
  \begin{array}{c}
    [\alpha - \beta \sin (2 \theta)] (i \omega S'^1 + \omega_L 
    S'^3) - i \omega \beta \cos( 2 \theta) S'^2\\  ( \alpha + \beta \sin
    (2 \theta) )  i \omega S'^2   -  \beta \cos ( 2 \theta) (i \omega
    S'^1 + \omega_L 
    S'^3)  \notag 
  \end{array}
\right) \, .
\end{align}
We consider the configuration $\theta = \theta' = - \pi/4$ such that
the SOI induced internal rf field is perpendicular to $\mathbf B_0$
and the longitudinal component $S'^2(t) = S_{\mathrm{eq}}$ is not
altered in linear response in $E$. Note that in this case
Eq.(\ref{eq:spin-current-RD}) simplifies such that $I^3_{x'} = \hbar(
i \omega S'^1 + \omega_L S'^3)/ (2 m (\alpha - \beta))$. This relation
differs from the naive model of an average spin-orbit field equating
the internal field $\boldsymbol{\Omega}(\mathbf p(t))$ with its
average $\boldsymbol{\Omega}(\langle\mathbf p \rangle (t))$.  Contrary
to Eq.(\ref{eq:spin-current-RD}), we then find $i \omega S'^1 +
\omega_L S'^3 = \Gamma^1 S'^1$ where $\Gamma^1$ is a phenomenological
transverse relaxation rate. Discrepancies to the model of an averaged
spin-orbit field occur similarly for other effects such as the
generation of an out-of plane polarization\cite{Engel2007} and
Zitterbewegung\cite{Schliemann2005}.

We proceed by evaluating the spin-Hall current $I^3_{x'}$ in terms of the
vertex correction Eq.(\ref{eq:vertex-correction}) which was obtained in the
diagrammatic approach and is valid up to second order in $a_R, a_D$. The
linear combination $i \omega S'^1 + \omega_L S'^3$ cancels in
lowest order (cf. Eq.(\ref{eq:sus3trans})) such that $I^3_{x'}$ is given
by the second order terms $q_{ij}, q$. From Eq.(\ref{eq:spin-pol-rd}) and
Eq.(\ref{eq:spin-current-RD}) we find the spin-Hall conductivity, defined as
$\sigma^{3, \mathrm{res}}_{x'y'} = \hbar I^3_{x'} /2 E(\omega)$, to be
given by 
\begin{align}
  \label{eq:spin-current-explicit} 
\sigma^{3, \mathrm{res}}_{x'y'} = \frac{e}{4 \pi} \frac{i \omega_L
\tau (\alpha^2 - \beta^2)}{(3 \alpha^2 - 2 \alpha \beta + 3 \beta^2) - i 2
\omega_L \tau (\alpha - \beta)^2}.
\end{align}
Remarkably, for high frequencies $\omega_L \tau (\alpha - \beta)^2 \gg
(\alpha + \beta)^2$ and $\alpha \neq \beta$
Eq.(\ref{eq:spin-current-explicit}) reaches the universal limit
$\sigma^{3, \mathrm{res}}_{x'y'} = |e|/8 \pi \times (\alpha +
\beta)/(\alpha - \beta)$ (independent of the disorder details). This
limit depends only on the ratio $\alpha/ \beta$ of the strengths of
the SOIs, but not on their absolute values, and agrees with the clean
limit found in\cite{Mishchenko2004} for $\beta = 0$. Indeed, for the
condition $\omega_L \tau \gg 1$ ($\omega_L=\omega$), many cycles of
the electric rf field pass through between subsequent scattering
events such that the system effectively behaves as ballistic. This
regime can be exploited to achieve high spin polarizations as
described above.  

Moreover, the singularity in
Eq.(\ref{eq:spin-current-RD}) for $\alpha = \beta$ is removed in
Eq.(\ref{eq:spin-current-explicit}) up to the accuracy $\mathcal
O(a_R^2, a_D^2, a_D a_R)$ considered here and we find that $\sigma^{3,
  \mathrm{res}}_{x'y'} $ vanishes in the configuration where $\bar
\chi$ is maximal, i.e. for $\alpha = \beta$ and $\theta = \theta' = -
\pi/4$.
 
We turn now to a geometrical interpretation of the spin Hall current relating
it to the trajectories $\boldsymbol{\mathcal{S}} = \{(S'^1(t),S'^3(t))| t \in
\mathbb{R} \}$ followed by the tip of the polarization vector. For an applied
electric field $\mathbf E(\omega) = E_0 \mathbf e_{||} [\delta(\omega -
\omega_0) + \delta(\omega + \omega_0)]/2 $ with frequency $\omega_0$ this
trajectory is given by the polarization (as a function of time)
\begin{align}
  \label{eq:spin-polarization}
\left(
  \begin{array}{c}
    S'^1(t) \\
    S'^3(t)
  \end{array}
\right)= \Lambda(\omega_0)
\left(
\begin{array}{c}
  \cos \omega_0 t \\
  \sin \omega_0 t
\end{array}
\right)
\end{align}       
with the matrix
\begin{align}
  \label{eq:a}
\Lambda(\omega_0) = E_{0}\left(
\begin{array}{cc}
   \mathrm{Re} \bar \chi^1 (\omega_0) &  - \mathrm{Im} \bar \chi^1(\omega_0) \\
   \mathrm{Re} \bar \chi^3 (\omega_0) &  - \mathrm{Im} \bar \chi^3(\omega_0) \\
\end{array}
\right) 
\end{align}
containing the Fourier components $\bar \chi^{1,3}(\omega)$ of the
susceptibility evaluated at $\omega = \omega_0$.  Eq.
(\ref{eq:spin-polarization}) constitutes a quadratic form for the trajectory
given by $\boldsymbol{\mathcal{S}} = \{(S'^1,S'^3) | \mathbf S'^{\mathrm t}
\cdot \Lambda_2 \mathbf S' =1\}$ with real, positive eigenvalues
$\lambda_{1,2}$ (say $\lambda_1 < \lambda_2$) of the defining matrix
$\Lambda_2 = (\Lambda^{-1})^{\mathrm t} \Lambda^{-1}$.  Thus,
$\boldsymbol{\mathcal{S}}$ is of elliptic shape with semi-major and semi-minor
axis $a = 1/\sqrt{\lambda_1}$ and $b = 1/\sqrt{\lambda_2}$, resp.  We can
further determine the angle $\delta$ enclosed by the semi-major axis of
$\boldsymbol{\mathcal{S}}$ and the $S'^1$ direction since the matrix
$\Lambda_2$ is diagonalized by a rotation $\delta$ around $S'^2$. The
polarization of Eq. (\ref{eq:spin-polarization}) can thus be written as
\begin{align}
  \label{eq:ellipse}
\left(
  \begin{array}{c}
    S'^1(t) \\
    S'^3(t)
  \end{array}
\right)  = \left(
  \begin{array}[c]{cc}
\cos \delta & -\sin \delta \\
\sin \delta & \cos \delta \\
  \end{array} \right)  \left(\begin{array}[c]{c}
a \cos(\omega_0 t + \varphi) \\
b \sin(\omega_0 t + \varphi)
\end{array} \right)\,.
\end{align}
Here, $\varphi$ is a phase shift between the electric field and the
polarization.  From Eqs.  (\ref{eq:spin-polarization}) and (\ref{eq:ellipse}),
we can relate the real and imaginary part of the susceptibilities $\bar
\chi^1$ and $\bar \chi^3$ to the parameters $a, b$, $\varphi$, and $\delta$.
In particular, we obtain the spin Hall current (Eq.(\ref{eq:spin-current-RD}))
at resonance ($\omega_L = \omega$) as
\begin{align}
  \label{eq:spin-current-ab-res}
  I^3_{x'}(\omega) = \frac{\hbar E(\omega) e^{i (\varphi - \delta)}}{2 m (\alpha-\beta)
    E_0} i \omega_L (a - b ) \, .
\end{align}
Eq. (\ref{eq:spin-current-ab-res}) provides a remarkable interpretation of the
spin Hall current in terms of the geometric properties of the orbit
$\boldsymbol{\mathcal{S}}$. The component $I^3_{x'}$ is given by a complex
phase depending on the rotation angle $\delta$ and the difference between the
semi-minor and semi-major axis $a -b$. In the linear response regime, the spin
Hall current characterizes the deviation from a circular orbit with $a = b$ to
an elliptic shape (with $a \neq b$). Therefore, $I^3_{x'}$ becomes accessible
in terms of simple geometric properties of $ \boldsymbol{\mathcal{S}} $ in
experiments capable of resolving individual polarization components.

In conclusion, we predict a substantially enhanced spin polarization
due to interference effects of Rashba and Dresselhaus SOI. The spin
Hall current associated with this polarization can be interpreted in terms
of the trajectory in spin space and vanishes if
the polarization is maximal.

We thank O. Chalaev, D.  Bulaev, J. Lehmann, and H.-A. Engel for
helpful discussions. We acknowledge financial support from the Swiss
NF, NCCR Nanoscience Basel, and the ONR.

\appendix
\section{Functions}\label{sec:functions}
For $\theta = -\pi/4$, $\omega = \omega_L$, and $\rho = \beta / \alpha$ the
functions $q$ and $q_{ij}$ are given by
  \begin{align}
    q_{11} &= \frac{- 2 a_R^2 (\lambda-1)^2 \lambda}{(1-2 \lambda)^2} \\ \notag
     & \times \left[ 1
      + 6 \lambda^2 (\rho -1)^2 + (\rho - 8) \rho - 4 \lambda (1 + (\rho -5)
      \rho) \right] \, ,
  \end{align}
  \begin{align}
    q_{12} &= \frac{- 2 a_R^2 (\lambda-1)^2 }{(1-2 \lambda)^2} \\ \notag
     & \times \left[ 2 \lambda^3 (\rho -1)^2 + 4 \lambda^2 \rho + (1 + \rho)^2
     - \lambda (3 + \rho (4 + 3 \rho))\right] \, ,
  \end{align}
  \begin{align}
    q_{13} &= \frac{i \sqrt{2}   a_R^2 (\lambda-1)^2 }{(1-2 \lambda)^2} \big[
    1 + 4 \lambda^3 (\rho -1)^2 + \rho (6 + \rho)  \\ \notag
     & - 4
       \lambda^2 (1 + (\rho -4) \rho) - 2 \lambda (1 + \rho (8 + \rho))\big] \, ,
  \end{align}
  \begin{align}
    q_{33} &= \frac{- 2   a_R^2 (\lambda-1)^2 }{\lambda (1-2 \lambda)^2}
    \big[4 \lambda^4 (\rho -1)^2 - 12 \lambda^2 \rho     \\ \notag
     & - (1 + \rho)^2 + 3 \lambda
    (1 + \rho)^2 - 4 \lambda^3 (1 + (\rho -4) \rho )\big] \, ,
  \end{align}
  and $ q_{22} = q_{11}$ and $ q_{23} = q_{13}$.


\end{document}